\documentclass[reprint,aip,twocolumn,groupedaddress,nobibnotes]{revtex4-1}

\usepackage{tabularx} 
 
\newcolumntype{A}{>{\centering\arraybackslash}X}

\usepackage{amssymb}
\usepackage{graphicx}
\usepackage[caption=false]{subfig} 
\usepackage{float}
\usepackage{amsmath}
\usepackage{dcolumn}
\usepackage{color}
\usepackage{graphicx}
\usepackage[pdftex,colorlinks=true,linkcolor=red,citecolor=blue]{hyperref}
\usepackage{soul}

\usepackage{multirow}
\usepackage{bigstrut}

\begin{document}

\title{The grain-size effect on thermal conductivity of uranium dioxide}
\author{K. Shrestha$^1$}
\author{T. Yao$^2$}
\author{J. Lian$^2$}
\author{D. Antonio$^1$}
\author{M. Sessim$^3$}
\author{M. Tonks$^3$}
\author{K. Gofryk$^1$}
\email{email: krzysztof.gofryk@inl.gov}
\affiliation{$^1$Idaho National Laboratory, Idaho Falls, Idaho, 83402, USA\\$^2$Department of Mechanical, Aerospace, and Nuclear Engineering, Rensselaer Polytechnic Institute, Troy, NY 12180, USA\\$^3$Department
Department of Materials Science and Engineering, University of Florida, Gainesville, FL 32611, USA}

\begin{abstract}We have investigated the grain boundary scattering effect on the thermal transport behavior of uranium dioxide (UO$_2$). The polycrystalline samples having different grain-sizes (0.125, 1.8, and 7.2 $\mu$m) have been prepared by spark plasma sintering technique and characterized by x-ray powder diffraction (XRD), scanning electron microscope (SEM), and Raman spectroscopy. The thermal transport properties (the thermal conductivity and thermoelectric power) have been measured in the temperature range 2-300~K and the results were analyzed in terms of various physical parameters contributing to the thermal conductivity in these materials in relation to grain-size. We show that thermal conductivity decreases systematically with lowering grain-size in the temperatures below 30 K, where the boundary scattering dominates the thermal transport. At higher temperatures more scattering processes are involved in the heat transport in these materials, making the analysis difficult. We determined the grain boundary Kapitza resistance that would result in the observed increase in thermal conductivity with grain size, and compared the value with Kapitza resistances calculated for UO$_2$ using molecular dynamics from the literature.

\end{abstract}


\maketitle

\section*{Introduction}
Uranium dioxide is one of the most studied actinide materials as it is used as the primary fuel in the commercial nuclear reactors.\cite{1,2,3} There are around 500 active nuclear reactors, producing more than 15\% of the total electricity worldwide. In a reactor, the heat energy produced from the nuclear fission events inside the fuel pellets is transformed into electricity. Thus, the heat transport mechanism, $i.e.$ thermal conductivity of the fuel material is an important parameter for fuel performance, regarding its efficiency and safety. A nuclear reactor operates at extreme environments that can include high temperature, high pressure, and high irradiation. As a result, a fuel pellet undergoes severe structural changes under irradiation conditions, including grain subdivision, fission gas bubbles growth and redistribution and extended defects accumulations.\cite{4,5} Thermal properties of the fuel material are greatly affected by these changes which ultimately affect the performance of a reactor. Numerous theoretical and experimental studies (see Refs.~\onlinecite{6,7,8,9} and references therein) have been carried out to understand how these microstructure changes affect thermal transport properties of UO$_2$.

UO$_2$ is a Mott-Hubbard insulator with an energy gap of $\sim$2~eV.\cite{10,11,12} It crystalizes in cubic, CaF$_{2}$ type of structure and orders antiferromagnetically at the N\'{e}el temperature, $T_{N}$ = 30.5 K.\cite{13,14} In an insulator, the lattice vibrations (phonons) responsible for the heat transport are scattered by different scattering centers, such as defects, grain boundaries, phonon-phonon, etc. Depending upon the temperature range, different scattering mechanisms dominate at different temperature regimes.\cite{15,16,17} For instance, umklapp phonon-phonon scattering dominates the thermal conductivity at high-temperature, while the point-defect and boundary scattering govern the heat transport at intermediate and low temperatures, respectively. At low temperatures where the phonon mean free path is comparable to the grain-size, the grain boundary scattering mechanism is the main factor limiting the thermal conductivity. The effect of grain-size on the thermal conductivity has been investigated at low temperatures in other types of materials, such as semiconductors, thermoelectrics, nanomaterials, and thin films.\cite{18,19,20,21,22} In the case of UO$_2$, most of the studies on thermal properties are focussed in the high temperature range (where nuclear reactors operate) to better understand the fuel performance and reactor design.\cite{23,24} However, in order to better understand mechanisms that govern heat transport in this important technological material, and to accurately model this compound at all relevant temperatures, the effects of various scattering mechanism must be quantified.

Here, we have carried out systematic studies on the grain-size effect on thermal conductivity of UO$_2$ by performing measurements at low temperatures to study different scattering mechanisms, focussing on grain boundary scattering. The UO$_2$ samples (having grain-sizes 0.125, 1.8, and 7.2 $\mu$m) have been synthesized by Spark Plasma Sintering technique and characterized by XRD, SEM, and Raman methods. We show that the grain boundary scattering parameters vary systematically with the grain-size below 30 K. Such a behavior is not observed at higher temperatures where other scattering processes start to dominate. The thermal conductivity data are analyzed using the Callaway model and the variation of different parameters with the grain-size are discussed. In addition, the grain boundary scattering has been assessed in these materials using molecular dynamic simulations at higher temperatures.

\begin{figure*}[t]
\centering
\includegraphics[width=1.0\linewidth]{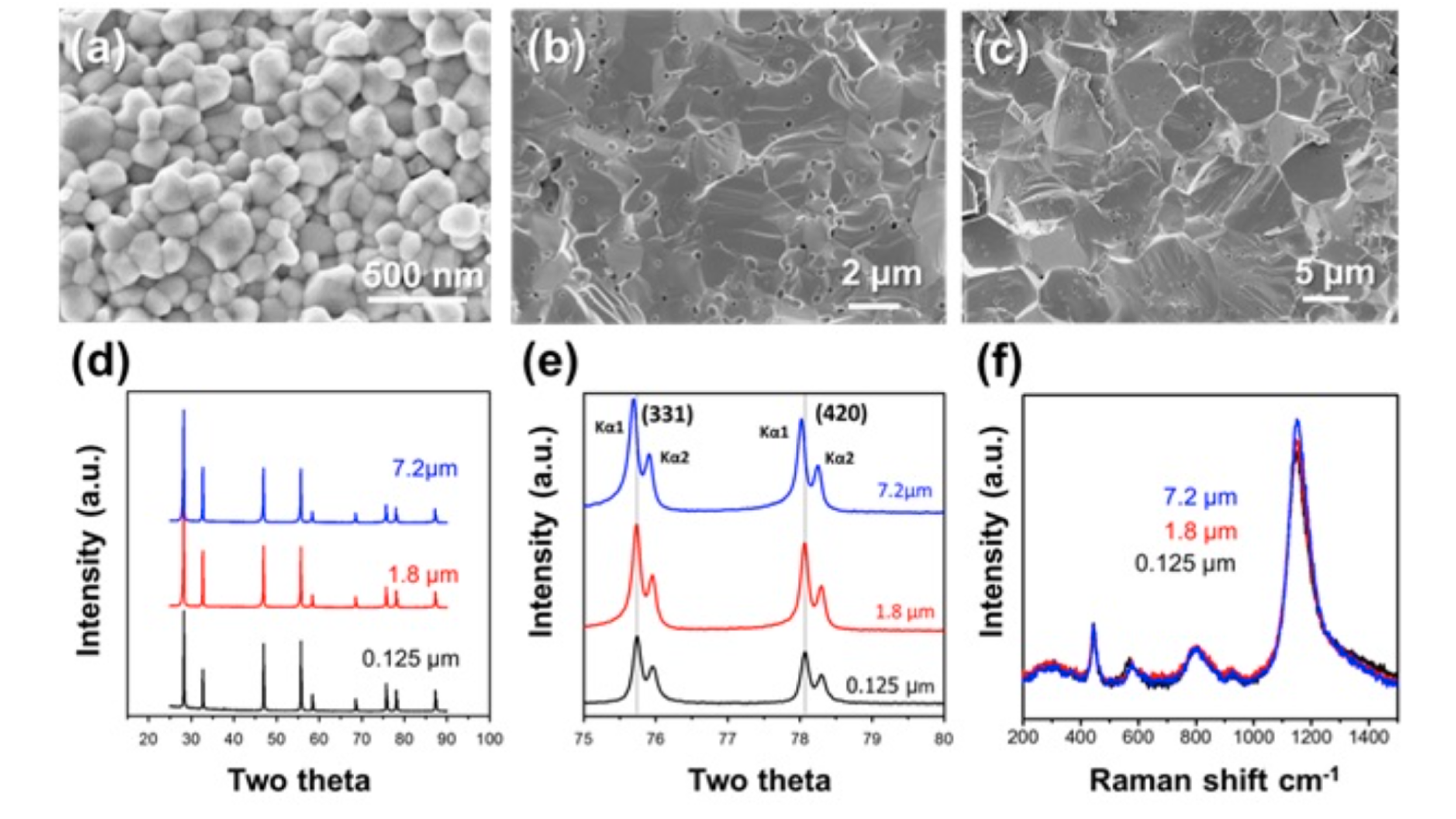}
\caption{(Color online) (a), (b), and (c) shows the microstructure features of sintered UO$_{2}$ fuel pellets with different grain-size of 0.125 $\mu$m, 1.8 $\mu$m, and 7.2 $\mu$m, respectively. (d) XRD spectra show that the sintered pellets have UO$_{2+x}$ structure with `x' values calculated by peak positions as shown in (e) for the high angle section. Superimposing feature of Raman spectra (f) indicates similar degree of interaction between defects and UO$_2$ crystal structure in the sintered pellets.}\label{XRD}
\end{figure*}

\section*{Experimental}
Polycrystalline UO$_2$ fuel pellets with three different grain-sizes (their physical properties are summarized in Table \ref{table1}) were sintered by spark plasma sintering from various batches of powder prepared from UO$_{2.16}$ powder purchased from International Bio-analytical Industries Inc., USA. Detailed information on powder samples can be found in Ref. [\onlinecite{25,26}]. Generally, the pellets with a grain-size of 7.2 $\mu$m were sintered directly from the as purchased UO$_{2.16}$ powder at 1600 $^o$C for 5 mins under a pressure of 40 MPa. The pellets with a grain-size of 1.8 $\mu$m were sintered from nano-crystalline UO$_{2.03}$ powder at 1300 $^o$C for 30 mins under a pressure of 40 MPa. Due to the graphite die used in those two sintering routes, the pellets were in-situ reduced to hypo-stoichiometric. The pellets with a grain-size of 0.125 $\mu$m were sintered at 700 $^o$C for 5 mins under a pressure of 500 MPa in WC die. These sintered pellets were hyper-stoichiometric and a post-sintering annealing was conducted in a tube furnace in 4\% H$_2$/Ar gas atmosphere in order to reduce oxygen. The furnace was purged by 4 hrs gas flow at a rate of 200 ml/min, then the reducing was conducted at 600 $^o$C for 24 hours at a gas flow rate of 50 ml/min. The sintered pellets are carefully stored in an oxygen controlled environment with momentary exposure to air for microstructure and phase characterization. The bulk density of the pellets was measured by an immersing method using DI water as the media, calculated based on weight difference in air and water, against a theoretical value of 10.97 g/cc for UO$_2$. Microstructure characterization was conducted using a Carl Zeiss Supra 55 (Jana, Germany) field emission SEM. Grain-size was determined using a rectangular intercept method following an ASTM E122-88 standard (1992). The average size is given by:

\begin{equation}\label{1}
D = \sqrt{\frac{4 A}{\pi \big(N_i + \frac{N_0}{2}\big)}}
\end{equation}
where A is the area of an arbitrary drawn rectangle, $N_i$ and $N_0$ are the numbers of grains in the rectangle and on the boundary of the rectangle, respectively. At least two hundred grains were analyzed for each pellet. The grain-size uncertainties are standard deviations of the measured grain-size for the same pellet from different locations.

\begin{table*}
\setlength{\tabcolsep}{20pt}
\caption{Physical properties of polycrystalline UO$_2$ samples.}
\centering
\begin{tabular}{l l l l l l l}
\hline\hline
Sample ID &UO$_2$ (0.125 $\mu$m)&UO$_2$ (1.8 $\mu$m)& UO$_2$ (7.2 $\mu$m)\\ [0.5ex]
\hline
Grain size ($\mu$m)&0.125$\pm$0.007&1.8$\pm$0.2&7.2$\pm$0.5\\
Stoichiometry&2.007$\pm$0.002&1.996$\pm$0.004&1.979$\pm$0.004\\
Theoretical density (\%)&96.5&95.8&96.2\\
\hline
\end{tabular}\label{table1}
\end{table*}

X-ray diffraction (XRD) spectra of the sintered pellets were collected by a Panalytical X$^{'}$Pert XRD system (Westborough, MA, USA) using Cu K$_{\alpha}$ ($\lambda$ = 1.5406 $\AA$) irradiation at room temperature. Before each run, the X-ray beam was aligned with a direct beam method through a 0.2 Cu beam attenuator. Sample height was aligned with respect to the X-ray beam using the bisect method. A scanning step of 0.013$^o$ with 2 seconds per step was used. The O/U ratio was determined from the following empirical equation: $a$ = 5.4705 $-$ 0.132$x$, \cite{27} where `$a$' is the derived lattice parameter and `$x$' is the stoichiometry deviation of UO$_{2+x}$ from stoichiometric UO$_2$. To estimate the O/U ratio, peaks in the region of 55 $-$ 90 were used as input and the calculated stoichiometries are statistically summarized. Micro-Raman spectra were collected at room temperature using a Renishaw Micro-Raman spectrometer excited by a green argon laser (514 nm). A typical spectrum was acquired with an exposure time of 10 seconds and 3 accumulations with a laser power of 20 mW. An extended scanning region from 200 to 1500 cm$^{-1}$ was chosen since it contains the featured peaks for UO$_2$. For each pellet, multiple locations were checked so that the spectrum is representative.

The thermal conductivity and Seebeck coefficient measurements of UO$_2$ samples were carried out in a Physical Properties Measurement System DynaCool-9  PPMS (Quantum Design) using the thermal transport (TTO) option and Pulse power method. The measurements were carried out using the continuous mode by slowly varying the temperature (0.2~K~min$^{-1}$). Typical dimensions of samples were $\sim$6$\times$1.2$\times$1.2~mm$^{3}$. 

\section{Results and Discussion}

Figs.\ref{XRD}a and c show the dense microstructure with various grain-sizes, as summarized in Table \ref{table1}. All pellets are fully densified with measured density higher than 95 \% TD. The XRD spectrum in Fig. {\ref{XRD}}d shows that the sintered pellets are single phase UO$_{2}$. Detailed spectra at the high angle area (Fig. {\ref{XRD}}e) shows well-separated K$_{\alpha_1}$ and K$_{\alpha_2}$ peaks for the (331) and (420) planes. The gray lines added sit on the exact two-theta angles for 0.125 $\mu$m samples. The peaks for 1.8 $\mu$m shifted slightly to lower angles, while the ones for 7.2 $\mu$m have a larger degree of peak shifting, indicating slight changes in the lattice parameter with stoichiometry. However, the superimposing features of the Raman spectra (Fig. {\ref{XRD}}f) shows the chemical bonding in those three-different grain-sized samples are very similar, indicating comparable localized defect interaction with the crystal structure of UO$_2$.

\begin{figure}[b]
\centering
\includegraphics[width=1.0\linewidth]{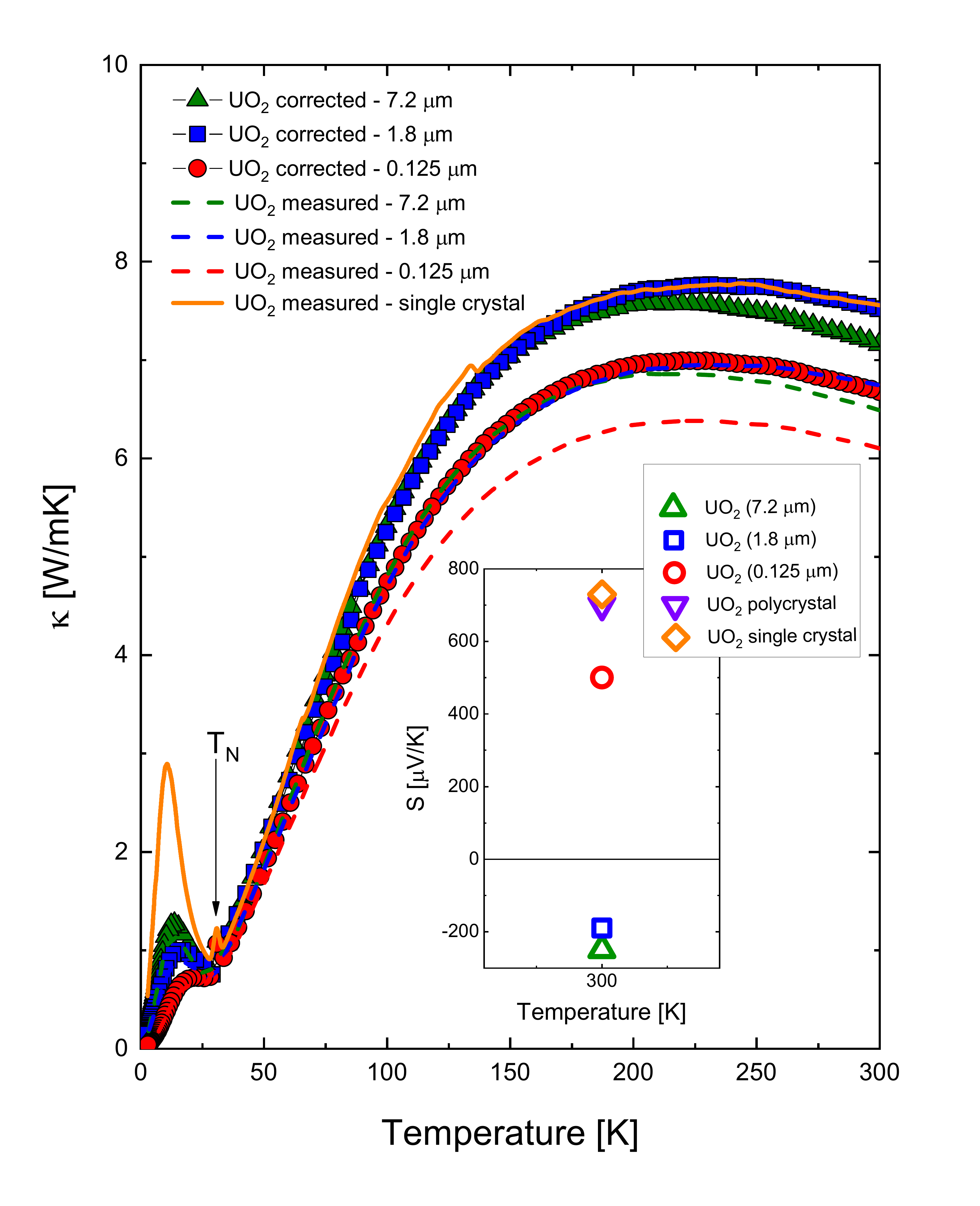}
\caption{(Color online) Temperature dependence of the thermal conductivity of the UO$_2$ polycrystalline samples having different grain-sizes. The dotted lines show the measured thermal conductivity of UO$_2$ polycrystals while the symbols represent the corrected values by taking into account the density difference (see the text). The solid orange line shows the UO$_2$ single crystal results (data taken from the Ref.~\onlinecite{28}). Inset: Seebeck coefficient of UO$_2$ samples measured at room temperature. The thermoelectric data for stoichiometric UO$_2$ polycrystalline sample is taken from Ref. \onlinecite{30}.}\label{Fig2}
\end{figure}

\begin{figure*}[t]
\centering
\includegraphics[width=1.0\linewidth]{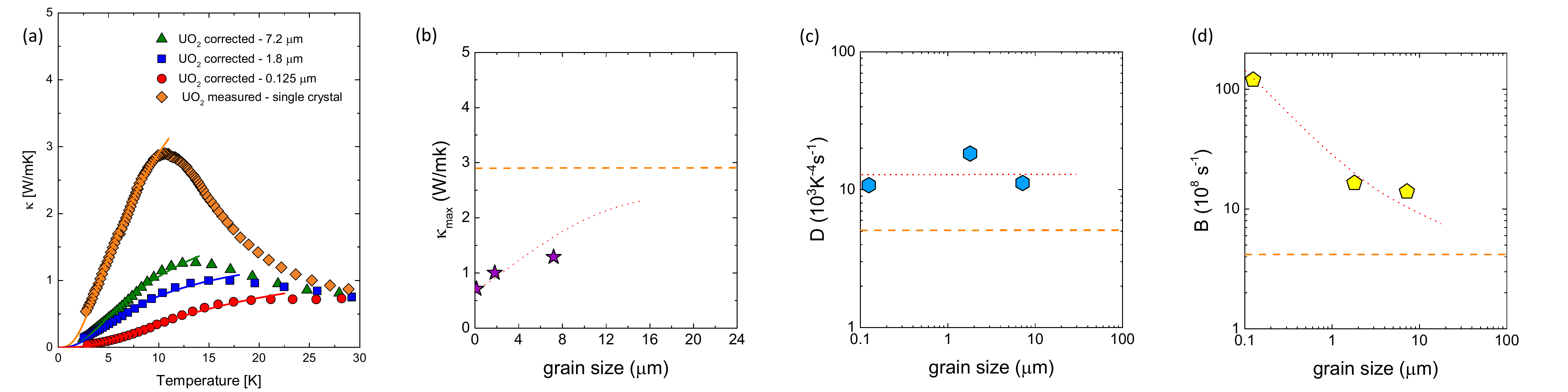}
\caption{(Color online) (a) The low temperature thermal conductivity of UO$_2$ samples. The solid lines represent the least-square fits of the Callaway model to the experimental data (see the text); the grain-size dependence of $\kappa_{max}$ (b), parameters $D$ (c), $B$ (d). The corresponding values for the UO$_2$ single-crystal are displayed as a dotted orange horizontal line in the relevant graphs. The red dotted lines are guides to the eye.}\label{Fig3}
\end{figure*}

Figure~\ref{Fig2} shows the thermal conductivity of UO$_2$ polycrystalline samples in the temperature range, 2-300 K. For comparison, we have also included the temperature dependence of the thermal conductivity of UO$_2$ single-crystal.\cite{28} The thermal conductivity of all samples shows similar temperature dependence as that of UO$_2$ samples in the previous reports.\cite{28,31,32} All $\kappa(T)$ curves consist a broad maximum at $T\sim$220 K and a minimum at the N\'{e}el temperature, $T_N$ = 30.5 K. In addition, there is a well-defined peak at $T\sim$10 K. Previous studies have revealed only a small difference in the thermal conductivity between single-crystal and polycrystalline UO$_2$ \cite{32a}. Above magnetic ordering, the thermal conductivity of uranium dioxide single crystal is limited by 3-phonon Umklapp scattering processes together with resonant scattering.\cite{28,32} These mechanisms are associated with a short mean free path, which may imply that grain boundaries have a relatively small effect on the phonon conduction at this temperature range (see Refs. \onlinecite{15,28,32}).

The measured thermal conductivities are all significantly lower than the single crystal value, even for the 7.2 $\mu$m grain size sample for which grain boundary scattering should be very low, as can be seen in the Figure~\ref{Fig2}. These $\kappa(T)$ values may be also compared to the one obtained for single crystal UO$_2$ material. Several possible scenarios besides grain boundary scattering may contribute to the reduction of the thermal conductivity in the polycrystalline samples as compared to a single crystal. The main source of the reduction comes from the fact that the polycrystalline UO$_2$ samples have slightly lower bulk densities ($\sim$96\%) than the single crystalline material (100\%).\cite{fink} In order to rescale our measured thermal conductivity values to 100\% density we have used the phenomenological expression proposed by Brandt and Neuer;\cite{bn} $\kappa_{0}=\kappa_{p}/(1-\alpha p)$, where $\alpha = 2.6 - 0.5t$. The parameter $p$ stands for a porosity factor, $t$ = T(K)/1000, and $\kappa_{0,p}$ is the thermal conductivity of fully dense ($p$ = 0) and porous UO$_2$, respectively. In Figure~\ref{Fig2} we also show the corrected thermal conductivity of our polycrystalline samples. After the density correction, the values of thermal conductivity are similar or slightly smaller than that of the single crystal. Another source that could impact the thermal conductivity might be related to the fact that the polycrystalline UO$_2$ samples were synthesized using natural uranium isotope, whereas the UO$_2$ single crystal consists of depleted uranium. Natural uranium contains slightly more fissile uranium U-235, about 0.72\%, then the depleted uranium, 0.2\%. This small percentage change of U-235 atoms, however, should have a negligible effect on the thermal conductivity value.\cite{34} Lastly, it has been shown that the oxygen off-stoichiometry, i.e. UO$_{2\pm x}$ has quite large impact on the thermal conductivity of uranium dioxide. Theoretical and experimental studies have shown that both hyper- and hypo-stoichiometry lower the thermal conductivity of UO$_2$, \cite{35,36,37,l1,l2} and its values decreases as much as 30\% for UO$_{2.033}$, as compared to stoichiometric UO$_2$ at room temperature.\cite{29} 

\renewcommand{\multirowsetup}{\centering}
\setlength{\tabcolsep}{20pt}
\begin{table*}
\centering
\caption{Comparison between measured thermal conductivity values with effective thermal conductivities calculated using Eq.\ref{kap1} with the new fit value for the Kaptiza resistance $R_{fit}$ and the value from molecular dynamics simulations from the literature $R_{MD}$. Though $R_{fit}$ is significantly larger than $R_{MD}$, the calculated thermal conductivities only differ for the smallest grain size.}
\begin{center}
\begin{tabular}{ccccc}
\hline\hline
\multirow{5}{2cm}{$d$ ($\mu$m)}& \multicolumn{3}{p{7cm}}{\centering Thermal Conductivity (W/mK)} \\ 
\cline{2-5} & \multicolumn{1}{c}{measured} &  \multicolumn{1}{c}{corrected}  & \multicolumn{1}{c}{Eq.\ref{kap1} with $R_{fit}$} & \multicolumn{1}{c}{Eq.\ref{kap1} with $R_{MD}$} \bigstrut \\ & \multicolumn{1}{c}{} & \multicolumn{1}{c}{} & \multicolumn{1}{c}{} & \multicolumn{1}{c}{} \\ 
\hline
single crystal &7.6 & - & - & -  \\
0.125 & 6.1	& 6.7	& 6.7 & 6.9  \\
1.8 & 6.8 &7.5 &7.5 & 7.5  \\
7.2	& 6.5 & 7.2 &	7.6	 & 7.6 \\
\hline
\end{tabular}
\end{center}
\end{table*}\label{t2}

In general, in order to precisely determine the oxygen content in UO$_2$ an x-ray diffraction is widely used. \cite{38} Here, we have also adapted thermopower measurements to probe the oxygen stoichiometry in this material. Depending upon the oxygen content, two types of charge carriers can exist in UO$_{2}$\cite{30,39}. In hyper-stoichiometric samples (UO$_{2+x}$) the positive hole-like carriers\cite{30} would lead to positive Seebeck effect, whereas, in hypo-stoichiometric UO$_{2-x}$ the negative electron-like carriers\cite{39} will result in negative Seebeck effect. The inset to Figure \ref{Fig2} shows the Seebeck coefficient ($S$) of the UO$_2$ samples measured at $T$ = 300 K. The Seebeck coefficient of the UO$_2$ single crystal is positive with the value $S\sim750\mu V/K$, which is also close to $S$ value for the polycrystalline UO$_2$.\cite{30} The $S$ value, however, changes for the different grain-size UO$_2$ samples and even changes sign for the samples having grain-size 1.8 and 7.2 $\mu$m. The negative sign of the Seebeck coefficient might suggest the presence of hypo-stoichiometric UO$_2$ in 1.8 and 7.2 $\mu$m samples. These results are consistent with the measured stoichiometries shown in Table \ref{table1}.


The presence of lower densities in the polycrystalline samples, as compared to single crystals, will reduce the thermal conductivity in these materials, especially at high temperatures. In addition, the variation of Seebeck coefficient and XRD measurements suggest that very small oxygen off-stoichiometry might be present and play a role in lowering of the thermal conductivity in the UO$_2$ samples. If so, this implies that separating different scattering mechanisms (especially grain boundary scattering) and sources of thermal resistance in UO$_2$ above room temperature is a challenging task. Grain boundary scattering dominates the thermal resistance of a material in the low-temperature regime when the grain-size is comparable or smaller than the mean free path of phonons.\cite{15} The grain boundaries behave then as scattering centers for phonons, which ultimately reduce the thermal conductivity values and governs the thermal conductivity in the low-temperature regime.\cite{15,16,17} Figure \ref{Fig3}a shows the blown-up region of the thermal conductivity curves of the UO$_2$ samples shown in Fig.\ref{Fig2} in the range below 30 K. As can been seen, in this temperature range the thermal conductivity decreases systematically with lowering the grain-size, as expected. The variation of thermal conductivity value $\kappa_{peak}$ measured at the peak position near $T$=10 K (see the arrows in Fig.\ref{Fig3}a) is shown in Fig. \ref{Fig3}b. It is observed that $\kappa_{peak}$ increases systematically with the grain-size as expected for grain boundary scattering and approaches towards the single crystal value (shown by the horizontal line) as the grain-size is increased. This dependence of thermal conductivity on the grain-size observed in UO$_2$ is consistent with grain boundary scattering being a main source of the thermal resistance.\cite{20,22}

In order to get more information of how different scattering processes affect the thermal conductivity in uranium dioxide, we have used the Callaway model\cite{40} to analyze the experimental data obtained. This phenomenological model has been previously used to successfully describe the lattice thermal conductivity in different materials.\cite{21,41,42,43} This approach takes into account scattering by different scattering mechanisms such as grain boundaries, point defects, or/and Umklapp phonon$-$phonon processes.\cite{40} At low temperatures (below $\sim$30~K), the thermal conductivity of insulators is mainly dominated by the boundary and point defects.\cite{19,44} Therefore, in order to analyze our low-temperature thermal conductivity data of UO$_2$, we have taken into account only the grain-boundary ($B$) and point-defect ($D$) scattering contributions. Within the framework of this model, the thermal conductivity can be expressed as,

\begin{equation}\label{Callaway}
\kappa(T) = \frac{k_B}{2\pi^2\it{v}}\bigg(\frac{k_BT}{\hbar}\bigg)^3\int_{0}^{\Theta_D/T}\frac{\tau_{ph}x^4e^x}{(e^x-1)^2}dx,
\end{equation}

where $v$ and $\tau_{ph}$ represent the mean velocity of sound and the phonon relaxation time, respectively. The parameter $x$ stands for $\hbar \omega/k_BT$. The parameters $\hbar$, $\Theta_D$, and $k_B$ are the reduced Planck's constant, the Debye temperature, and Boltzmann constant, respectively. The mean sound velocity was determined using the formula $v = k_B\Theta_D/{\hbar\sqrt[3]{6\pi^2n}}$, where $n$ is the number of atoms per unit volume. Taking $\Theta_D$ = 395 K\cite{45}, $v$ is estimated to be 3,171 m s$^{-1}$ for UO$_2$. The relaxation time is taken as the sum of inverse relaxation times of the scattering processes, i.e. $\tau_{ph}^{-1} = \tau_D^{-1} + \tau_B^{-1}$. The particular inverse relaxation times are given by the following expressions:

\begin{equation}\label{Defect}
\tau_D^{-1} = Dx^4T^4 = D\bigg(\frac{\hbar\omega}{k_B}\bigg)^4
\end{equation}

and

\begin{equation}\label{Boundary}
\tau_B^{-1} = B,
\end{equation}

where $D$ and $B$ are the fitting parameters. The $B$ value is large for the lower grain-size sample and it decreases while increasing the grain-size as expected for grain boundary effect.\cite{20,22}

The solid lines in Fig.~\ref{Fig3}a represent fits of the Callaway model to the low temperature data of UO$_2$. The variations of the obtained parameters with grain-size are shown in Figs.~\ref{Fig3}c and d. The results for UO$_2$ single crystal (no grain boundary scattering) has been shown as a dashed orange horizontal line in the corresponding graphs. The grain boundary effect, described by the parameter $B$, is higher in the smaller grain-size sample, and it decreases towards the value for the UO$_2$ single-crystal as the grain-size is increased. The parameter $D$, related to defect scattering, is comparable to each other (see Fig. \ref{Fig2}c) suggesting the presence of similar number of defect scattering centers in the measured samples. 

An alternative means of quantifying the impact of grain boundary scattering on the thermal conductivity is the grain boundary (GB) Kapitza resistance (see Refs. \onlinecite{k1,k2,k3} and references therein). The GB Kapitza resistance $R$ can be calculated using the following equation, $R=d/\kappa_{eff}-d/\kappa_{sc}$, where $\kappa_{sc}$ is the single crystal thermal conductivity and $\kappa_{eff}$ is the effective thermal conductivity of a polycrystal of grain size $d$. By solving for the effective thermal conductivity, we obtain the equation:

\begin{equation}\label{kap1}
\kappa_{eff} = d\frac{\kappa_{sc}}{R \kappa_{sc}+d}
\end{equation}

Molecular dynamics simulations have been used to calculate the Kapitza resistance in various UO$_2$ GB's (see Table 1 in Ref.~\onlinecite{mt1}). The largest Kapitza resistance found was $R_{MD}$ = 1.69$\times$10$^{-9}$ m$^{2}$K/W at a temperature of 300 K. The data obtained in this work provides an excellent means of calculating the Kapitza resistance from experimental data by determining the value for $R$ that results in an effective thermal conductivity using Eq.\ref{kap1} that is closest to the corrected values for the three different grain sizes. Using this approach, $R_{fit} = 2.30\times10^{-9}$ m$^2$K/W at 300 K; the effective thermal conductivity values using this value with the grain sizes from the three samples are shown in Table II and have a maximum error (compared to the corrected values) of 5.7\%. If the uncertainty in the grain size measurements reported in Table I are considered, the standard deviation of $R$ at 300 K is found to be $0.13\times10^{-9}$ m$^2$K/W. The $R_{fit}$ value from the experiments is 4.7 standard deviations larger than the value from the molecular dynamics simulations, indicating that the difference can not be explained with just experimental error. It is not surprising that the experimental value is larger than the molecular dynamics value, since the simulations assume a perfectly stoichiometric grain boundary with no impurities and will thus have less scattering. However, the molecular dynamics value was close enough to the experiential value that it did not add significantly more error in the calculated effective thermal conductivities, as shown on Table II.

\section*{Conclusion}
To summarize, we have synthesized UO$_2$ samples having different grain size (0.125, 1.8, and 7.2 $\mu$m) and investigated the grain-size effect on thermal properties in this material. The samples have been characterized by x-ray powder diffraction (XRD), scanning electron microscope (SEM), and Raman spectroscopy. By performing low-temperature thermal conductivity measurements we have studied the grain-boundary scattering related to grain-size and its impact on the thermal conductivity in these materials. Although the operating temperatures in nuclear reactors are high ($\sim$1000 K), many important physical characteristics such as the effect defects and grain boundary scattering on the heat transport are all emphasized at moderate or low temperatures. At high temperature various different scattering mechanisms are present simultaneously, making their separation and detailed analyses difficult. By performing measurements at low temperatures (below 30 K), where the grain boundary and defect scatterings dominate the thermal transport, we show a systematic dependence of the thermal conductivity on the grain-size. Such a behavior is not observed at higher temperatures due to other scattering processes that govern thermal resistance of these materials. In particular, (i) the porosity has a large impact on the thermal conductivity of all polycrystalline samples, (ii) grain boundary scattering has a large impact on the sample with a small grain size but little impact for the two larger grain size samples, (iii) the presence of U-235 likely has no significant impact, and (iv) the stoichiometry could have had some impact. The measured thermal conductivities were also used to determine the Kapitza resistance of UO$_2$ at 300 K, and the value was significantly larger than a value from the literature obtained using molecular dynamics simulations. The knowledge of the details of the grain boundary scattering mechanisms in UO$_2$ will be useful for researchers working on modeling and simulations of this nuclear fuel. The approach presented here would be also useful to study thermal transport in other applied materials, especially thermoelectrics.

\section*{acknowledgements}
The work was supported by the DOE's NE Nuclear Energy University Programs (NEUP), US Department of Energy's Early Career Research Program, and Advanced Fuel Campaign.


\end{document}